# Localisation And Imaging Methods for Moving Target Ghost Imaging Radar Based On Correlation Intensity Weighting

Yuliang Li,
Shanghai Institute of Optics and Fine Mechanics,
Chinese Academy of Sciences (CAS),
Shanghai, People's Republic of China,
Email: lyl931206@163.com

*Abstract*—Ghost imaging radar is a new system of gaze imaging radar with high detection sensitivity, super-resolution and better anti-interference performance, but the relative motion between the radar system and the target will make the target imaging deteriorate. This paper proposes to perform absolute position localisation of a single target in the field of view by weighting the correlation strength of a single frame image of rough target, and to compensate translation of the reference arm speckle according to the localisation and tracking trajectory to accumulate the rough image into a high quality image. The proposed correlation intensity weighted localization and tracking imaging method has been verified by simulation to be able to locate and image targets in the field of view well.

*Keywords—ghost imaging, rough image, intensity weighting, speckle translation*

## I. Introduction

In contrast to conventional LIDAR, ghost imaging radar is a type of gaze imaging radar based on photon rise and fall properties, which has advantages such as high detection sensitivity, super-resolution and interference resistance [1-5]. However, the relative motion between the radar system and the target will lead to poor correlation between the reference arm speckle and the bucket detection observations, and poor or even unimaginable imaging quality. Common solutions include fast sampling with increased frame rate and scatter motion compensation schemes [6], the former of which is very demanding in terms of hardware requirements and therefore more research has been conducted based on the latter.

Research on correlated imaging and localisation of moving targets has focused on the tangential motion of the target relative to the system. The research process is generally based on the assumption of quasi-static estimation [7], i.e. the tangential motion process is studied in segments, where the target can be considered as stationary and within the field of view with respect to the imaging system during each segment. In [8-9], researchers obtained motion pattern estimates based on a motion state search conditioned on the optimal quality of the correlated reconstructed images, which required repeated image reconstruction to determine the correct motion parameters. In [10-12], some scholars used the method of solving the maximum value of the correlation function of two rough correlated images to obtain relative position pose estimates of the target, which could not recover the tracking of the target in case of transient recovery of tracking in case of loss. Reference [13] divided the scatter into different sub-scatters based on the variation of object arm barrel detection sampling data and the motion of the target throughout the illumination pattern, and determined the target position by finding the corresponding overlap between the target barrel detection value and some parts of the scatter. Reference [14] used the barrel detector intensity rise and fall signal associated with object spatial information for blood cell counting, applying motion target ghost imaging to the biomedical field to achieve high throughput blood cell classification and counting. The literature [15] combines laser tracking with conventional four-quadrant detectors, replacing the barrel detector with a four-quadrant detector to achieve real-time localization of the Specklewhile correlating the four-quadrant light intensity summation with the reference arm scattering to obtain the target image, but the method cannot give a position estimate for small targets that lie entirely in one quadrant. Other papers [16-17] use structured light to illuminate the target area, exploit the natural optical convolution properties, and use barrel detection reception to detect and identify the relative offset of the target, but the nature of the imaging is not the same as random scatter-based correlation imaging.

In this paper, we propose a method for moving target localization and imaging of ghost imaging radar based on correlation intensity weighting, in which the target coordinate position is obtained by intensity weighting of individual pixel coordinates based on one undersampled rough image of the target in a simple background, and the speckles panned to reconstruct the target at the same position in the field of view by combining the obtained target position, and a high-quality target image is obtained by superimposing the reconstructed multiple rough images. This method is computationally simple and gives a high quality image of the target. It can obtain the absolute coordinates of the target under the detection field of view, avoid the disadvantage that the relative position offset detection cannot be retraced, and can achieve better localization and correlation imaging for a single target in the field of view.

## II. Principle of Ghost Imaging and Intensity-Weighted Localisation

### A. Principle of Ghost Imaging

The principle of ghost imaging is shown in Fig.1, where a laser is passed through a rotating glass to produce a controlled light intensity rise and fall of the thermal light source. One light travels a distance and is sampled by a CCD to obtain a series of speckled light intensity distributions $I_i(x,y)$ (where $i=1\cdots K$, $K$ is the number of samples taken when reconstructing the target, and $I_i(x,y)$ is the light intensity at the point (x,y) on the scatter) to become the reference arm. The other way of light irradiation is transmitted or reflected on the target surface to the spatially non-resolvable barrel detector PMT, which collects the total light field intensity $y_i$, where it can be considered that $y_i$ is the information obtained by the irradiated scattered spots encoding or modulated by the target.

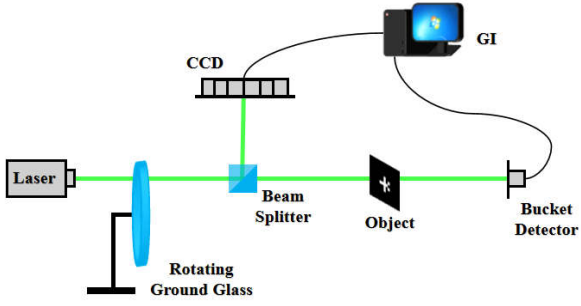

Fig. 1. Ghost imaging schematic

The target object is obtained by performing a second order correlation operation between the light intensity distribution matrix obtained from the reference arm CCD and the total intensity value obtained from the object arm barrel detector:

$$O_{GI}(x,y) = \frac{1}{K}\sum_{i=1}^{K}(I_i(x,y) - \langle I_i(x,y)\rangle)(y_i - \langle y_i\rangle) \quad (1)$$

Where, $\langle \bullet \rangle$ denotes the overall average.

*B. Object Localization with Rough Image*

In this paper, the target is first coarsely imaged $O_{rGI}$ by a small amount of undersampling. According to (1), the intensity of each pixel $(x,y)$ in the imaged space can be considered as the cumulative average of K correlation operations, and the intensity of the target region in the reconstructed rough image is significantly stronger than that of the non-target region. In other words, the scatter of the reference arm at the target region is associated with the barrel detection value with a significant correlation strength, while the non-target region has a weak correlation calculation, and the stronger the correlation strength the greater the probability that it will become a pixel in the target region. The intensity feature of a rough image is the target feature. In this paper, the coordinates of regions with strong association strength values are extracted and weighted normalized to their coordinates to obtain an estimate of the absolute coordinates of the target.

Before estimating the target centre coordinates, a pixel correlation intensity filter is required. Because the intensity characteristics of the background region are weak, a more accurate target pixel can be obtained by filtering the pixels with larger amplitude to filter out the background region:

$$g(x^s, y^s) = \frac{O_{rGI}(x,y)}{\max O_{rGI}(x,y)} \geq t \quad s.t. \ (x^s, y^s) \in \Omega \quad (2)$$

Where, $g(x,y)$ is the normalised association strength distribution, $t \in [0,1]$ is an artificially set screening threshold, $(x^s, y^s)$ is the coordinates of the pixels that satisfy the condition, and $\Omega$ is the set of pixels that satisfy the condition.

The intensity values of the filtered rough images are normalised and weighted to obtain the intensity weighting function:

$$W(x^s, y^s) = \frac{g(x^s, y^s)}{\sum_{\Omega} g(x^s, y^s)} \quad (3)$$

Multiply the weighting function with all the pixel coordinates that satisfy the condition and sum to get the final target position estimate $(x_c, y_c)$:

$$\begin{cases} x_c = \sum_{\Omega} x^s \cdot W(x^s, y^s) \\ y_c = \sum_{\Omega} y^s \cdot W(x^s, y^s) \end{cases} \quad (4)$$

*C. Reference Arm Speckle Translation and Rough Image Stacking to Optimize Image*

Based on the estimated coordinates obtained, the reference arm speckle can then be translated. In order to ensure that the position of the reconstructed target relative to the field of view range remains unchanged, the speckle is translated to the origin position and the new speckle obtained by the translation is expressed as:

$$I_{inew}(x,y) = I_i(x - x_c, y - y_c) \quad (5)$$

Where $I_{inew}(x,y)$ is the reference arm speckle matrix after translation and the distance of the translation is the calculated target centre coordinates. Reconstruction of the target image requires that the spots at different locations are again correlated with the reference arm barrel detections after translation to calculate:

$$\begin{aligned} O_{GInew}(x,y) &= \frac{1}{N}\sum_{i=1}^{N}(I_{inew}(x,y) - \langle I_{inew}(x,y)\rangle)(y_i - \langle y_i\rangle) \\ &= \frac{1}{r}\sum_{j=1}^{r} O_{jnewGI}(x,y) \end{aligned} \quad (6)$$

Where $N = rK$ is the total number of samples, $r$ is the number of sheets of rough image, $K$ is the number of samples required to generate a rough image, and the final reconstructed target image is equivalent to the summed and averaged result of $r$ sheets of reconstructed rough image after the speckle translation transformation. Using a rough image that requires only a small number of samples for moving target localisation, the effective information of the target area is gradually accumulated as the moving target moves, and a clear image is eventually obtained.

III. SIMULATION AND ANALYSIS

In order to verify the effectiveness of the proposed correlation intensity weighted localisation and imaging method, simulations of its localisation and imaging are carried out. The effect of the screening threshold t and the number of samples K of the generated rough image on the localisation and imaging is analysed in this paper. The simulation process of target motion follows the trajectory of the target-fitted trigonometric curve, with a reference arm speckle and reconstructed image field-of-view size of 64x64 pixels and a target size of 15x15 pixels, and a 2x2 pixels Bernoulli speckle used for the reference arm spot. The target moves once per K samples in the field of view. In this paper, the positioning accuracy and imaging quality are evaluated using the position root-mean-square error (PRMSE) and the peak signal-to-noise ratio (PSNR).

$$PRMSE = \frac{1}{r}\sum_{j=1}^{r}\sqrt{(x_{jc}^{o}-x_{jc})^{2}+(y_{jc}^{o}-y_{jc})^{2}} \quad (7)$$

$$MSE = \frac{1}{mn}\sum_{x=1}^{m}\sum_{y=1}^{n}\left\|O_{GInew}(x,y)-O_{original}(x,y)\right\|^{2} \quad (8)$$

$$PSNR = 20\cdot\lg(O_{GInew\max}/\sqrt{MSE}) \quad (9)$$

where, $x_{jc}^{o}$ and $y_{jc}^{o}$ are the original target centre position, $x_{jc}$ and $y_{jc}$ are the calculated target centre position, $O_{original}(x,y)$ is the original target image, $O_{GInew}(x,y)$ is the reconstructed image, and MSE is the image signal-to-noise ratio.

At first, this paper simulates and analyzes the screening threshold, given the number of samples K=300, and takes the mean square error of target localization under different screening thresholds, and plots the relationship between the screening threshold t and localization accuracy in Fig.2 When larger the number of samples of strongly correlated points obtained by screening becomes smaller, the influence of a few mutated strongly correlated points becomes larger, and the centre of localisation deviates from the original target centre instead; when smaller more background noise points are selected as localisation weights, making the localisation error large.

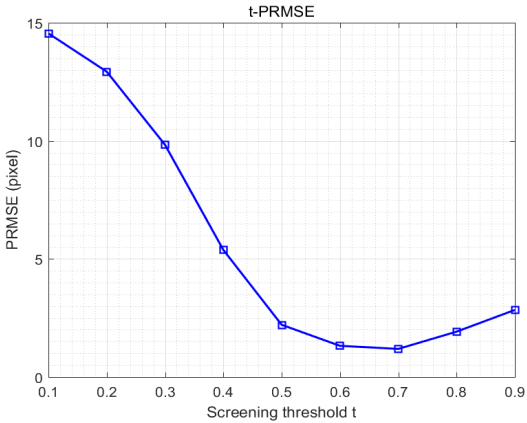

Fig. 2. Relationship graph of screening threshold *t* and positioning accuracy

In this paper, we also investigate the influence of the number of undersampling K on the localization of the generated rough image, and select the screening threshold t=0.7. The relationship between the number of samples and the positioning accuracy is plotted in Fig.3. As the number of samples increases, the positioning accuracy improves significantly, but too high a number of samples makes the time interval between each target positioning larger and the probability of the target flying out of the field of view increases, which is not conducive to tracking.

The target image reconstruction effect is shown in Figure 4 and the target trajectory is shown in Figure 5 for the case of setting K=300 and t=0.7. Fig.4a shows the original target image, Fig.4b shows the ghost image without speckle shift compensation, Fig.4c shows the rough image and Fig.4d shows the final superimposed high quality image. The PSNR of the final high quality image is 21.25 *dB* and the PSNR of the rough image is 15.07 *dB*. The peak signal-to-noise ratio of the target image obtained by the superimposed method is significantly improved.

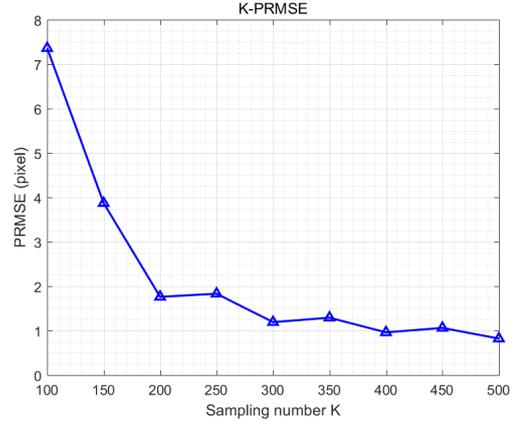

Fig. 3. Relationship graph of undersampling *K* and positioning accuracy

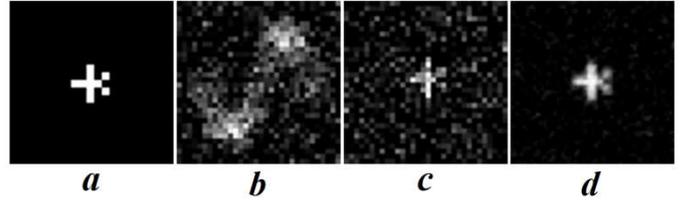

Fig. 4. Comparison of ghost imaging reconstruction results

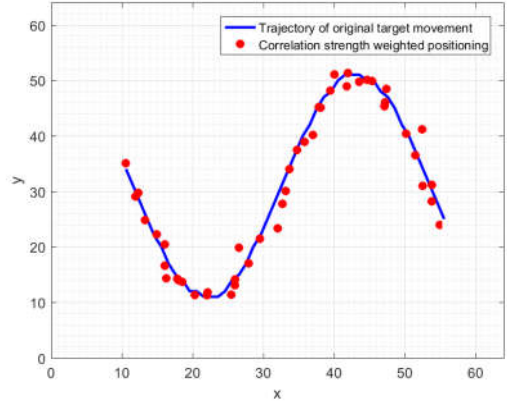

Fig. 5. Positioning trajectory map

## IV. CONCLUSION

This paper proposes a method to obtain the absolute position of a single target in the field of view by weighting the correlation intensity of a single frame of rough image, compensating the reference arm speckle translation according to the localization and tracking trajectory, and finally accumulating the rough image into a high quality image. This method is simple to calculate and can obtain the absolute coordinates of the target under the detection field of view, avoiding the disadvantage that the relative position detection cannot be retraced, and can achieve a good localization and ghost imaging of the moving single target in the field of view.